\baselineskip=18pt
\def\a{\alpha}\def\b{\beta}\def\d{\delta}\def\e{\epsilon}
\def\f{\phi}
\def\m{\mu}\def\n{\nu}\def\r{\rho}\def\s{\sigma}

\def\D{\Delta}\def\G{\Gamma}\def\L{\Lambda}
\def\O{\Omega}\def\U{\Upsilon}

\def\na{\nabla}
\def\inf{\infty}\def\mo{{-1}}\def\ha{{1\over 2}}

\def\mn{{\mu\nu}}

\def\af{asymptotically flat }
\def\fe{field equations }\def\bh{black hole }\def\as{asymptotically }

\def\ssy{spherically symmetric }
\def\dr{dimensionally reduced }

\def\sch{Schwarzschild }\def\ads{anti-de Sitter }

\def\KK{Kaluza-Klein }

\def\GR{general relativity }
\def\GB{Gauss-Bonnet }

\def\ab{asymptotic behavior }

\def\section#1{\bigskip\noindent{\bf#1}\smallskip}

\def\PL#1{Phys.\ Lett.\ {\bf#1}}

\def\PR#1{Phys.\ Rev.\ {\bf#1}}\def\CQG#1{Class.\ Quantum Grav.\ {\bf#1}}
\def\NP#1{Nucl.\ Phys.\ {\bf#1}}
\def\JMP#1{J.\ Math.\ Phys.\ {\bf#1}}

\def\PRep#1{Phys.\ Rep.\ {\bf#1}}

\def\ref#1{\medskip\everypar={\hangindent 2\parindent}#1}
\def\beginref{\begingroup
\bigskip
\centerline{\bf References}
\nobreak\noindent}

\def\er{{\cal R}}\def\es{{\cal S}}\def\eg{{\cal G}}\def\ef{e^{2\f}}


\def\er{{\cal R}}\def\es{{\cal S}}\def\eg{{\cal G}}\def\ef{e^{2\f}}

{\nopagenumbers
\line{\hfil January 2007}
\vskip80pt
\centerline{\bf Asymptotic properties of black hole solutions}
\centerline{\bf in dimensionally reduced Einstein-Gauss-Bonnet gravity}
\vskip40pt
\centerline{
{\bf S. Mignemi}\footnote{$^\ddagger$}{e-mail:
smignemi@unica.it}}
\vskip10pt
\centerline {Dipartimento di Matematica, Universit\`a di Cagliari}
\centerline{viale Merello 92, 09123 Cagliari, Italy}
\centerline{and INFN, Sezione di Cagliari}
\vskip100pt
\centerline{\bf Abstract}
\vskip10pt
{\noindent
We study the \ab of the \ssy solutions of the system obtained
from the dimensional reduction of the six-dimensional Einstein-\GB action.
We show that in general the scalar field that parametrizes the size of the internal space
is not trivial, but nevertheless the solutions depend on a single parameter.
In analogy with other models containing \GB terms,
naked singularities are avoided if a minimal radius for the horizon is assumed.}
\vskip100pt\
P.A.C.S. Numbers: 04.70.Bw 04.50.+h
\vfil\eject}

\section{1. Introduction}

In order to explore the physical implications of the dimensional reduction of higher-dimensional
models of gravity, it is interesting to investigate the properties of the solutions of the Einstein equations
having the form of the direct product of a four-dimensional black-hole spacetime with an internal space
of microscopic size. In the simplest case, the dimensionally reduced theory will consist in four-dimensional
general relativity coupled to a scalar field which parametrizes the size of the internal space.
The global properties of the solutions of this form have been investigated some time ago in [1],
where it was shown that all solutions having an internal space of finite size are \af for pure Einstein
gravity, or \as \ads if a cosmological constant is present.

However, in the context of higher-dimensional theories of gravity, higher-curvature corrections to the
Einstein-Hilbert action arise naturally. Among these, a special role is played by the \GB (GB) terms [2].
These are in fact the only possible generalization to higher dimensions of the EH action that gives rise to
\fe which are second order in the metric, linear in the second derivatives, and divergence free.
These corrections also appear in the low-energy effective action of string theories.
Of course, the presence of additional terms in the action modifies the properties of the \bh solutions
of the dimensionally reduced theory.

In two recent papers [3-4], we have classified the black hole solutions of a dimensionally reduced
six-dimensional Einstein-GB model, with and without the inclusion of a cosmological constant. We have
shown that all \ssy solutions with \as constant scalar field have either flat or \ads behavior at infinity
(in addition to these, de Sitter-like solutions exist with no asymptotic region).
However, we have not studied the details of the solutions. In view of the no-hair theorems [5] of Einstein
gravity, it would be interesting to investigate more thoroughly the behavior of the solutions at infinity, to
see if scalar hair can be present when a GB term is added. It is known, in fact, that in the case of
GB-scalar couplings in four dimensions, a non-trivial scalar field can arise, whose scalar charge is however
not an independent parameter, but a function of the \bh mass [6].\footnote{*}{When multiple scalar fields are present, independent scalar charge can however appear [7].}

In the present paper, we study this problem by expanding the solutions of the \fe at infinity in powers of
${1\over r}$, and show that for \bh solutions the scalar field configuration is always non-trivial, even if
the scalar charge vanishes due to the rapid decrease of the field at infinity.

We also investigate the near-horizon behavior of the solutions. It is known that in the case of GB black holes
one must impose a minimal value to the horizon radius $r_h$ in order to prevent naked singularities [8-9].
By expanding the solutions near the horizon in powers of $r-r_h$, we show that
the same phenomenon occurs in our dimensionally reduced model.
The near-horizon expansion also permits to check the possibility of degenerate horizons in the de Sitter-like
case, which are related to a generalization of the Nariai metric.

\bigskip

We consider a six-dimensional model with action
$$I=\int\sqrt{-g}\ d^6x\ (2\b+\er+\a\es),\eqno(1)$$
where $\er$ is the curvature scalar, $\es=\er^{\m\n\r\s}\er_{\m\n\r\s}-4\er^{\m\n}\er_{\m\n}+\er^2$
the quadratic GB term (which is the only relevant one in six dimensions), $2\b$ the cosmological constant,
and $\a$ a coupling constant.

We assume that the metric has the form of the direct product of a \ssy four-dimensional spacetime
with a maximally symmetric two-dimensional space. Hence we can write the metric ansatz in its
six-dimensional form as
$$ds^2=-\D dt^2+\D^\mo dr^2+R^2d\O_e^2+S^2d\O_i^2,\eqno(2)$$
where $d\O^2_i$ is the line element of the maximally symmetric two-dimensional space and $\D$, $R$
and $S$ are functions of $r$.

The field equations then read
$$\eqalignno{&\D'RR'S^2+\D R'^2S^2+\D'R^2SS'-\D R^2S'^2-4\D RR'SS'+\e R^2-S^2-\b R^2S^2=\cr
&4\a[1-\e(\D'R'R-\D R'^2)-(\D'S'S-\D S'^2)+3\D R'S'(\D'R S+\D R S'+\D R'S)],&(3)\cr
&\cr
&S^2(\D''R+2\D'R'+2\D R'')+4S(\D'RS'+\D R'S'+\D RS'')+2\D RS'^2-2\e R-2\b RS^2=\cr
&-4\a[\e(\D''R+2\D'R'+2\D R'')-2S(\D'^2R'S'+\D\D''R'S'+\D\D'R''S'+\D\D'R'S'')\cr
&-S'^2(\D'^2R+\D\D''R+6\D\D'R'+2\D^2R'')-S'S''(2\D\D'R+4\D^2R')],&(4)\cr
&\cr
&R^2(\D''S+2\D'S'+2\D S'')+4R(\D'SR'+\D S'R'+\D SR'')+2\D SR'^2-2S-2\b SR^2=
\cr&-4\a[\D''S+2\D'S'+2\D S''-2R(\D'^2S'R'+\D\D''S'R'+\D\D'S''R'+\D\D'S'R'')\cr
&-R'^2(\D'^2S+\D\D''S+6\D\D'S'+2\D^2S'')-R'R''(2\D\D'S+4\D^2S')],&(5)}$$
where $\e=1$ if the curvature of the internal space is positive, $\e=-1$ if it is negative, and $\e=0$ if it
vanishes.

The ground state of the theory is assumed to be the direct product of a
four-dimensional maximally symmetric spacetime of curvature $\L_e$ with a two-dimensional maximally symmetric
internal space of curvature $\L_i$. With this ansatz, the \fe reduce to
$$12\a\L_e^2+6\L_e+\b=0,\qquad12\a\L_e\L_i+\L_i+3\L_e+\b=0,\eqno(6)$$
with solutions
$$\L_e=-{6\pm\G\over24\a},\qquad\L_i=-{6-8\a\b\pm(1-4\a\b)\G\over8\a(5-12\a\b)},\eqno(7)$$
where $\G=\sqrt{36-48\a\b}$.
Eqs.\ (6) have two different solutions which, depending on the values of the parameters $\a$ and $\b$,
can have different nature. If $\a>0$, the first solution is $adS^4\times H^2$ for any $\b<{3\over4\a}$;
the second solution is $dS^4\times S^2$ if $\b<0$, $M^4\times R^2$ if $\b=0$, $adS^4\times H^2$ if $0<\b<{5\over12\a}$, $adS^4\times S^2$ if ${5\over12\a}<\b<{3\over4\a}$, $adS^4\times R^2$ if $\b={3\over4\a}$.
No solutions exist if $\b>{3\over4\a}$. We denoted as $M^4$, $dS^4$, $adS^4$ four-dimensional Minkowski,
de Sitter and anti-de Sitter spacetime, respectively, and by $R^2$, $S^2$ and $H^2$ two-dimensional maximally symmetric spaces of zero, positive and negative curvature, respectively.

If $\a<0$, the first solution is $dS^4\times S^2$ for any $\b>{3\over4\a}$; the second
solution is $adS^4\times H^2$ if $\b>0$, $M^4\times R^2$ if $\b=0$, $dS^4\times S^2$ if ${5\over12\a}<\b<0$,
$dS^4\times H^2$ if ${3\over4\a}<\b<{5\over12\a}$, $dS^4\times R^2$ if $\b={3\over4\a}$. No solutions exist
if $\b<{3\over4\a}$. In the following, we shall mainly be concerned with the case of \ads or flat spacetime,
since they have an asymptotic region, and hence for simplicity will consider only the case of positive $\a>0$,
which is also that dictated by string theory. Our results can however easily be extended to $\a<0$.

The action (1) can be dimensionally reduced, using the \KK ansatz
$$ds^2=-e^{-2\f}ds_e^2+\ef ds_i^2,\eqno(8)$$
where $\f$ is a scalar field parametrizing the size of the internal space, $ds_e^2$ is the line element of the four-dimensional
spacetime and $ds_i^2$ that of the two-dimensional internal space.
In particular, the spacetime metric can be written in the \sch form
$$ds_e^2=-Adt^2+A^\mo dx^2+B^2d\O_e^2,\eqno(9)$$
where $\ef=S^2$, $A=\D S^2$, $B=RS^2$, and a new radial variable $x=\int S^2dr$ has been defined.

The \dr action reads
$$\eqalignno{I=\int&\sqrt{-g}\ d^4x\Big[(1+2\a\e\,e^{-2\f})\er+\a\,e^{2\f}\es+
8\a\,e^{2\f}\eg_\mn\na^\m\phi\na^\n\phi&\cr
&+4\left(1+3\a\e\,e^{-2\f}\right)(\na\f)^2-8\a\, e^{2\f}(\na\f)^4+2\b\,e^{-2\f}+\e\,e^{-4\f}
\Big],&(10)}$$
Where $\er$, $\es$ and $\eg_\mn$ are the curvature, the GB scalar and the Einstein tensor of the
four-dimensional spacetime. The action (10) displays some similarity to the effective string action studied
in [6,8,9], and contains several non-standard couplings of the scalar field with gravity, that rule out the
usual no-hair theorems. The \fe are however still second order.

\section{2. Asymptotic behavior}
We first investigate the \ab of the solutions of eqs.\ (3)-(5), by expanding them in powers of ${1\over r}$.
Since the special case of vanishing cosmological constant cannot be obtained as a limit of
the general case, we start by considering this.
\bigskip
{\bf A. $\b=0$}
\medskip
As shown in [3], when  $\b=0$ the solutions can be either asymptotically flat, with flat internal
space, or \as anti-de Sitter, with internal space of negative curvature.
In the \af case, we expand the metric functions for $r\to\inf$ as
$$\D=\sum_{i=0}^\inf{\d_i\over r^i},\qquad R=\sum_{i=0}^\inf{\r_i\over r^{i-1}},
\qquad S=S_0\sum_{i=0}^\inf{\s_i\over r^i}.\eqno(11)$$
Obviously, one can put $\s_0=1$. Moreover, one is free to linearly transform the coordinate $r$.
This freedom can be fixed choosing $\r_0=1$, $\r_1=0$.

Substituting in the field equations (3)-(5), with $\e=0$, and comparing order by order in $1\over r$,
one obtains that $\d_0=1$, while $S_0$ and $\d_1$ are free parameters. Moreover,
$\d_2=\d_3=\r_2=\r_3=\s_1=\s_2=\s_3=0$. Up to this order, the \ab is
that of the \sch solution of dimensionally reduced pure Einstein gravity. The first
corrections due to the GB term arise at the fourth order:
$$\d_4={\a\over2}\,\d_1^2,\qquad\r_4={5\a\over12}\,\d_1^2,\qquad\s_4=-{\a\over4}\,\d_1^2.\eqno(12)$$

Putting $\d_1=-2M$, the metric fields read
$$\D=1-{2M\over r}+{2\a M^2\over r^4}+\dots,\qquad R=r\left(1+{5\a M^2\over3r^4}+\dots\right),
\qquad S=S_0\left(1-{\a M^2\over r^4}+\dots\right).\eqno(13)$$
In terms of the four-dimensional metric functions (9), after a rescaling of $x$,
$$A=1+\d_1\,{S_0\over x}+(\d_4+2\s_4)\,{S_0^4\over x^4}+\dots,\qquad B=x\left[1+\left(\r_4+{\s_4\over3}
\right)\,{S_0^4\over x^4}+\dots\right],$$
$$\ef=S_0^2\left[1+2\s_4\,{S_0^4\over x^4}+\dots\right],\eqno(14)$$
or, more explicitly,
$$A=1-{2m\over x}+\dots,\qquad B=x\left(1+{4\a m^2S_0^2\over3x^4}\right)+\dots,\qquad
\ef=S_0\left(1-{2\a m^2S_0^2\over x^4}+\dots\right),$$
where we have defined $m=-S_0\d_1/2$.
Note that the fourth order correction to $A$ vanishes. The four-dimensional metric is a deformation of the
\sch solution,
with nontrivial scalar field. The solution depends only on one parameter (the mass) and the scalar charge
vanishes due to the rapid decrease of the scalar field at infinity.
\medskip
In the \as\ads case, the expansion of the metric fields at infinity must be modified to take into
account the terms proportional to $r^2$ in the metric function $\D$. We put
$$\D=\sum_{i=0}^\inf{\d_i\over r^{i-2}},\qquad R=\sum_{i=0}^\inf{\r_i\over r^{i-1}},
\qquad S=S_0\sum_{i=0}^\inf{\s_i\over r^i}.\eqno(15)$$

The zeroth order equation, with $\e=-1$, gives the background solution
$$\d_0={1\over2\a},\qquad S_0=\sqrt{10\a\over3}.\eqno(16)$$
Choosing as before the gauge $\r_0=1$, $\r_1=0$, $\s_1=1$, one
obtains $\d_2=1$, $\d_3$ free, $\d_1=\d_4=\d_5=0$, $\r_i=0$ for
$2\le i\le5$, $\s_i=0$, for $1\le i\le5$. In this case, the
corrections driven by the GB term arise at sixth order:
$$\d_6=-{7\a\over10}\,\d_3^2,\qquad\r_6=-{7\a^2\over10}\,\d_3^2,\qquad\s_6={\a^2\over10}\,\d_3^2.\eqno(17)$$
Putting $\d_3=-2M$, the metric functions read
$$\D={r^2\over2\a}+1-{2M\over r}-{14\a M^2\over5r^4}+\dots,\qquad
R=r\left(1-{14\a^2M^2\over5r^6}+\dots\right),$$
$$S=\sqrt{10\a\over3}\left(1+{2\a^2M^2\over5r^6}+\dots\right).\eqno(18)$$

In terms of the four-dimensional metric functions (9), after a rescaling of $x$,
$$A=\d_0\,{x^2\over S_0^2}+1+\d_3\,{S_0\over x}+\left(\d_6+{6\d_0\s_6\over5}\right)\,{S_0^4\over x^4}+\dots,
\qquad B=x\left[1+\left(\r_6+{3\s_6\over5}\right)\,{S_0^6\over x^6}+\dots\right],$$
$$\ef=S_0^2\left[1+2\s_6\,{S_0^6\over x^6}+\dots\right],\eqno(19)$$
or, more explicitly,
$$A={3x^2\over20\a^2}+1-{2m\over x}-{128\a^2m^2\over15x^4}\dots,\quad B=x\left(1-{536\a^4m^2\over9x^6}\right)
+\dots,\quad\ef={10\a\over3}\left(1+{80\a^4m^2\over9x^6}+\dots\right),$$
where $m=-S_0\d_3/2$.

In this case the solution is a deformation of the Schwarzschild--\ads metric with nontrivial scalar field, but
again depends on a single parameter and exhibits no scalar charge.

\bigskip
{\bf B. $\b\ne0$}
\medskip
We pass now to discuss the general case $\b\ne 0$ [4]. Solutions with asymptotic regions exist
whenever $\a\b\le{3\over4}$, and are always \as anti-de Sitter. More precisely, there are two
possible \ads solutions if $\b>0$ and only one if $\b<0$. The internal space has negative
curvature for both solutions if $0<\b<{5\over12\a}$, or positive for one solution
and negative for the other if $\b>{5\over12\a}$ (if $\b={5\over12\a}$ there exists a unique
solution, with negative curvature).
If $\b<0$, the internal space of the \ads
solution has positive curvature. Finally, when $\b={3\over4\a}$, the internal space is flat.

In view of the \ads behavior of the solutions under consideration, one can expand as in (15).
At zero order one gets
$$\d_0={6\pm\G\over24\a},\qquad S_0=\sqrt{\left|{6-8\a\b\mp(1-4\a\b)\G\over2\b(3-4\a\b)}\right|}\,,
\eqno(20)$$
where $\G=\sqrt{36-48\a\b}$.

Expanding, with the usual gauge
conditions, one obtains $\d_2=1$, $\d_3$ free, $\d_1=\d_4=\d_5=0$,
$\r_i=0$ for $2\le i\le5$, $\s_i=0$, for $1\le i\le5$. The first
nontrivial terms arise again at sixth order:
$$\d_6=-{7\a(3-4\a\b)(30-24\a\b\pm5\G)\over10(3+2\a\b)[30-40\a\b\pm(5-4\a\b)\G]}\,\d_3^2,
\qquad\r_6=-{42\a^2(3-4\a\b)\over5(3+2\a\b)(6-8\a\b\pm\G)}\,\d_3^2,$$
$$\s_6=\pm{3\a^2(1-4\a\b)\G\over(3+2\a\b)[30-40\a\b\pm(5-4\a\b)\G]}\,\d_3^2.\eqno(21)$$
The four-dimensional metric can still be obtained using equations (19).

The asymptotic properties of these solutions are analogous to those of the \ads branch of the $\b=0$ case.
Actually, the limit $\b\to0$ diverges for the lower sign, and reduces to the \ads solution of the previous
section for the upper sign.
\bigskip
{\bf C. $\b={3\over4\a}$}
\medskip
When $\b={3\over4\a}$, the solution (20) diverges. One must
therefore consider this case separately. It turns out that the internal space is flat.
At order zero,
$\d_0={1\over4\a}$, while $S_0$ is undetermined.
Expanding, with the usual gauge conditions, one obtains $\d_1=\d_4=\d_5=0$, $\d_2=1$, $\d_3$ free,
$\r_i=0$ for $2\le i\le5$, $\s_2=\s_4=\s_5=0$, $\s_3$ free. The first
nontrivial terms in the expansion are
$$\d_6=-{2\s_3(3\s_3+8\a\d_3)\over 5\a}\qquad\r_6=-{4\s_3(3\s_3-2\a\d_3)\over 5},
\qquad\s_6=-{2\a\d_3(3\s_3+\a\d_3)\over3},\eqno(22)$$
In this case, a new free parameter $\s_3$ appears in the solutions.
Putting $\d_3=-2M$, $\s_3=D$
$$\D={r^2\over4\a}+1-{2M\over r}-{2D(3D-16\a M)\over 5\a r^4}+\dots,
\qquad R=r\left(1-{4D(3D+4M)\over 5r^6}+\dots\right),$$
$$S=S_0\left(1+{D\over r^3}+{4\a M(3D+\a M)\over3r^6}+\dots\right).\eqno(23)$$

The solution is again a deformation of the Schwarzschild--\ads metric, but depends on two parameters.
However, the scalar charge vanishes, since the field $S$ decreases as $r^{-3}$ at infinity.
\bigskip
{\bf D. $\b={1\over4\a}$}
\medskip
This is an algebraic special case, since the \fe are not sufficient to determine univocally the
solution (an analogous situation for \ssy solutions has been discussed in [10]).
With the ansatz $S^2=4\a$, and $ \e=-1$, the solution is given by an arbitrary function $R(r)$,
with $\D(r)$ satisfying the differential equation
$$\ha(R^2-4\a\D R'^2+4\a)\D''+2(RR'-2\a R'R''\D')\D'-2\a R'^2\D'^2+2(R'^2+2RR''+{R^2\over2\sqrt\a})\D=1+{R^2\over4\a}.$$
A special solution was found in [11] for $R=r$.

\section{3. Near-horizon behavior}
Also the behavior of the solutions near the horizon can be studied perturbatively. In models containing
GB terms, the request of absence of naked singularities imposes a lower value for the horizon radius [8,9].
In order to investigate this topic in the present model, we expand the metric functions near the horizon
at $r=r_h$ as
$$\D=L'_h(r-r_h)+\ha L''_h(r-r_h)^2+\dots,\qquad R=R_h+R'_h(r-r_h)+\dots,\qquad S=S_h+S'_h(r-r_h)+\dots,
\eqno(24)$$
with constant coefficients $L'_h$, $R_h$, $S_h$, etc.
Again we use the freedom of linearly transforming $r$ to put $R_h=r_h$, $R'_h=1$.
Then the \fe yield, in the limit $r\to r_h$
$$\eqalignno{&r_hS_h^2L''_h+4S_hS'_hL'_h+2S_h^2L'_h-2\e r_h-2\b r_hS_h^2+
4\a[2\e(r_hL''_h+L'_h)-2 S_hS'_hL'^2_h-r_hS'^2_hL'^2_h]=0,&(25)\cr
&r_h^2S_hL''_h+2r_h^2S'_hL'_h+4r_hS_hL'_h-2S_h-2\b r_h^2S_h+
4\a(S_hL''_h+2S'_hL'_h-S_hL'^2_h-2r_hS'_hL'^2_h)=0,&(26)\cr
&r_hS_h^2L'_h+r_h^2S_hS'_hL'_h-\e r_h^2-S_h^2-\b r_h^2S_h^2+4\a[S_hS'_hL'_h+\e(r_hL'_h-1)]=0.&(27)}$$
From (27) one can obtain $L'_h$ in terms of $S'_h$, $S_h$ and $r_h$:
$$L'_h={\e(4\a+r_h^2)+S_h^2(1+\b r_h^2)\over r_h(4\e\a+S_h^2)+S_hS'_h(4\a+r_h^2)}.\eqno(28)$$
One can then eliminate $L''_h$ from (25) and (26), and substitute (28).
In this way, one obtains a complicated second degree algebraic equation for $S'_h$ in terms of
$r_h$ and $S_h$. The equation admits real solutions for $S'_h$ if its discriminant $\U$ is positive.
We write it for $S_h=1$:
$$\eqalign{\U=&[1-4\b\a+4(\b^2-2\e\b^2-12)\a^2+(16\e\b^2+96\b-144\e)\a^3+(64\b^2+384\e\b)\a^4]r_h^6\cr
&-8\a^2[(\b-2\e)-2(\b^2-2\e\b+3)\a+8(\e\b^2-2\b+3\e)\a^2-48\e\b\a^3]r_h^4\cr
&-16\a^2[3-3(2\b+\e)\a-4(\b^2+2\e\b+31)\a^2+48(2\b-5\e)\a^3]r_h^2\cr
&-64\a^3[4-3(2\b-\e)\a-12(2\e\b+5)\a^2-144\e\a^3].}\eqno(29)$$
The roots of $\U$ depend on the specific values of the coupling constants $\a$ and $\b$ and their expression
is awkward.
However, in general, $\U\ge0$ for $r_h\ge\bar r_h$, where $\bar r_h$ is a function of $\a$, $\b$ and $S_h$.
Hence, a minimal value of the horizon radius is found, in analogy with other GB models.
In particular, a simple result is obtained in the special case $\b={1\over4\a}$, where $r_h^2\ge12\a$.

Another interesting possibility is that $L'_h$ can vanish if $\e$, $\a$ or $\b$ is negative. In this
case one has a double zero of the metric function, indicating  a degenerate horizon. This happens whenever
$$S_h^2=-\e\,{r_h^2+4\a\over\b r_h^2+1}.\eqno(30)$$
One can check that this possibility can only arise in the case of de Sitter-like black holes, when $\b<0$.
In this case,
$$r_h^2=S_h^2=-{1\over\b}\left(1+\sqrt{1-4\a\b}\right).\eqno(31)$$
With these values of the parameters, one should obtain a generalization of the Nariai metric.
In fact, they correspond to the exact solution $dS^2\times S^2\times S^2$,
$$ds^2=-\left(1-{r^2\over r_h^2}\right)dt^2+\left(1-{r^2\over r_h^2}\right)^\mo dr^2+r_h^2 d\O_e^2+r_h^2 d\O_i^2,$$
with $r_h^2$ given by (31).
\bigbreak
\section{4. Conclusion}
We have studied the \ab of the \bh solutions of \dr six-dimensional Einstein-GB theory. They can be either
\af or \as \ads and depend on a single parameter (the black hole mass), except in the special case
$\b={3\over4\a}$ , where a second parameter is present. The scalar field is always non-trivial for
non-vanishing mass, but its scalar charge is zero, due to its rapid decrease at infinity.
The situation is therefore analogous to that of the four-dimensional Einstein-GB model coupled to a scalar
field [6].
It must also be remarked that in the case $\b={1\over4\a}$ the solution is not completely determined by the field
equations.

The study of the near-horizon behavior of the solutions shows that also in this case a minimal value for the
horizon radius must be imposed in order to avoid naked singularities, as for other GB models [8,9]. Black
holes without asymptotic regions, analogous to Schwarzschild--de Sitter black holes of \GR, can develop a
degenerate horizon, leading to a generalization of the Nariai metric.

\beginref
\ref [1] S. Mignemi and D.L. Wiltshire, \CQG{6}, 987 (1989);
D.L. Wiltshire, \PR{D44}, 1100 (1991).
\ref [2] D. Lovelock, \JMP{12}, 498 (1971);
B. Zwiebach, \PL{B156}, 315 (1985);
B. Zumino, \PRep{137}, 109 (1986).
\ref [3] S. Mignemi, \PR{D74}, 124008 (2006).
\ref [4] M. Melis and S. Mignemi, \PR{D75}, 024042 (2007).
\ref [5] J.D. Bekenstein, \PR{D5}, 1239 (1972); \PR{D5}, 2403 (1972).
\ref [6] S. Mignemi and N.R. Stewart, \PR{D47}, 5259 (1993).
\ref [7] S. Alexeyev and S. Mignemi, \CQG{18}, 4165 (2001).
\ref [8] P. Kanti,  N.E. Mavromatos, J. Rizos, K. Tamvakis and E. Winstanley, \PR{D54}, 5049 (1996).
\ref [9] T.  Torii,  H.  Yajima, and K. Maeda, \PR{D55}, 739 (1997);
S.O. Alexeyev and M.V. Pomazanov, \PR{D55}, 2110 (1997).
\ref [10] J.T. Wheeler, \NP{B273}, 732 (1986).
\ref [11] H. Maeda and N. Dadhich, \PR{D74}, 021501 (2006).
\end